\documentclass[preprint,prb,aps,showpacs,superscriptaddress]{revtex4}
\usepackage{graphicx}
\def\be{\begin{equation}}
\def\ee{\end{equation}}
\def\bea{\begin{eqnarray}}
\def\eea{\end{eqnarray}}
\def\bdm{\begin{displaymath}}
\def\edm{\end{displaymath}}
\def\ba{\begin{array}}
\def\ea{\end{array}}

\begin{document}

\title{Spin current in topologically trivial and nontrivial noncentrosymmetric superconductors}

\author{Chi-Ken~Lu\footnote{
Present e-mail: chikenl@sfu.ca}} \affiliation{Institute of
Physics, Academia Sinica, Nankang, Taipei 115, Taiwan}
\affiliation{Department of Physics, Simon Fraser University, 8888
University Drive, Burnaby, B.C. V5A 1S6, Canada}\author{Sungkit
Yip} \affiliation{Institute of Physics, Academia Sinica, Nankang,
Taipei 115, Taiwan}

\date{\today }

\begin{abstract}

We study theoretically the surface of time-reversal-symmetric,
noncentrosymmetric superconductor with mixed singlet and triplet
order parameters.  A pair of counterpropagating subgap
quasiparticle surface bound states with opposite spin projections
are obtained in the nontrivial Z$_2$ case where the triplet
component is larger than the singlet one, contributing to a spin
current with out-of-plane spin projection. In contrast to the pure
$p$-wave cases, these subgap states do not have fixed spin
projections, which however depend on the momenta along the
surface. In the trivial Z$_2$ case where the singlet order
parameter is larger, no subgap surface bound states show up.  In
both cases, there is also a finite contribution to the spin
current from the continuum states with energies between the two
gaps. The method for obtaining the quasiclassical Green's
functions associated with the noncentrosymmetric superconductors
is also presented.

\end{abstract}


\maketitle

\section{Introduction}

How to manipulate spin in condensed-matter system has been the
main challenge for both experimentalists and theorists in this
community. Recently, the predictions\cite{Bernevig,QSH,TI3D} and
observations\cite{QSH_QW,TIexp,Bi2Se3,Shen} of topological
insulators (TI) with time-reversal symmetry(TRS), such as
HgTe/(Hg,Ce)Te semiconductor wells,\cite{QSH_QW}
Bi$_{1-x}$Sb$_x$,\cite{TIexp} Bi$_2$Se$_3$,\cite{Bi2Se3} and
Bi$_2$Te$_3$,\cite{Shen} inspire a great deal of interest in both
the application and the fundamental research ends . The main
feature of a TI is a pair of counter-propagating edge states with
opposite spin projections developed out of a gapped band
structure, very much like the edge states in the integer Quantum
Hall case except that the TRS is broken in the latter case. In the
language of homotopy,
the mapping from the momentum space to the Hamiltonian can be
smoothly deformed into either one of the two distinct elements in
the so-called Z$_2$ class associated with the trivial and
nontrivial TI's. This pair of surface bound states contribute to a
spin current near the surface of a TI.\cite{TIrev}


The concept of TI can also be generalized to superconductors (SC).
A simple example is the (2D) $p$-wave superconductor with its
order parameter given by $\vec{d}=k_x\hat{y}-k_y\hat{x}$. We can
see explicitly from the order parameter $\propto
[(k_x-ik_y)|\uparrow\uparrow\rangle+(k_x+ik_y)|\downarrow\downarrow\rangle]$
that the Cooper pairs with down(up) spins have a
counterclockwise(clockwise) motion. This state is clearly
time-reversal symmetric.  Each species is in an axial state,
considered in, e.g., Ref.\cite{ReadGreen,helicalSC}, with the
phase of the Cooper pair wavefunction advances (decreases) by $2
\pi$ when the angle of the momentum direction advances by the same
angle. Considering the surface of this superconductor adjacent to
vacuum, the incident and reflected quasiparticles see an order
parameter with a different phase factor, analogous to the case of
a Josephson junction.  A pair of surface bound states with
opposite spin projections propagate in opposite directions,
generating no charge but finite spin current, as in the case of a
topological insulator. Moreover, straightforward generalization of
the results of Ref.\cite{ReadGreen,helicalSC} implies that a
singly quantized vortex of this superconductor possesses a pair of
zero-energy Majorana states inside its vortex core. In contrast,
the order parameter of an $s$-wave singlet superconductor do not
change sign over the Fermi surface. No surface bound states are
topologically required and the superconductor thus belongs to
trivial class.

For noncentrosymmetric superconductors (NcSC), such as the
compounds CePt$_3$Si,\cite{Bauer-r} Li$_2$Pt$_3$B,\cite{exp_Li}
and CeRhSi$_3$,\cite{Ce113} and the 2D electron gas between two
insulating layers,\cite{Reyren} complication arises due to the
presence of parity-broken spin-orbital interaction and the
singlet-triplet mixed superconducting order
parameters.\cite{Gorkov} For example, for a system with no up-down
reflection symmetry, the s-wave singlet order parameter $\Delta_s$
and the triplet order parameters $\Delta_p$ with the d-vector
given by $(k_x\hat{y}-k_y\hat{x}$) would naturally mix since they
are of the \emph{same} symmetry.\cite{YipGarg}  Alternatively, the
two Fermi surfaces with opposite helicities can be associated with
two different superconducting gaps as a result of the broken
inversion symmetry.\cite{YipPRB2002} Nevertheless, the topological
classifications still can be deduced from the existence of
zero-energy vortex bound states\cite{Lu2} or the surface bound
states.\cite{Tanaka} These studies reveal that only the relative
signs of the pairing terms on the opposite-helicity bands matter
for the topology, which can also be shown by more explicit
topological arguments.\cite{YipTopology,ZhangTopology} NcSC with
opposite signs of pairing on the two bands resembles the pure
$p$-wave triplet superconductor in that a pair of topologically
protected zero energy or surface bound states reside within the
vortex core or at the surface, respectively, while these
topological bound states do not exist at all in the NcSC with same
sign of pairing, which resembles the pure $s$-wave singlet SC.


In this paper we shall consider in more detail the surface of a
NcSC, as shown in Fig.\ref{fig1}, as a function of the singlet
$\Delta_s$ and triplet $\Delta_p$ order parameters. For
simplicity, here we would not consider the dispersion along the z
axis (or effectively 2D) nor the splitting of the fermi surfaces
due to spin orbit interactions. The absence of spin-orbital
coupling may in fact correspond to a narrowly accessible physical
regime in mixed-parity superconductivity.\cite{Sergienko} Our
simplification however does not affect the main physics that we
would like to explain in this paper. Comments on this aspect will
be given later.

For the present geometry, there is no xy plane reflection
symmetry, whereas the time-reversal symmetry and the xz plane
reflection symmetry are left intact. Following the same symmetry
arguments as in our previous paper,\cite{Lu3}  the magnetization
along any directions automatically vanish due to the time-reversal
invariance. However, the spin currents $J^z_y$, $J^x_y$, and
$J^y_x$ can in principle be allowed.  We shall examine whether
these spin currents are finite.

In the pure p-wave case ($|\Delta_s|$=0), the bound states have
their spins quantized parallel or antiparallel to the z-axes. Only
$J^z_y$ is non-zero.\cite{Eschrig,Lu3}  We would examine what
happen to these states when $\Delta_s$ is finite, especially their
spin directions. We shall see that the spins are no longer
polarized along $z$ in the general case. In the context of
topological superconductor, the parity-mixed order parameters with
$|\Delta_p|>|\Delta_s|$ belong to the nontrivial Z$_2$
class,\cite{YipTopology,Tanaka,Lu2} and the pair of surface bound
states with opposite spin projections propagating in opposite
directions still exist. These surface bound states can in
principle be detected by tunneling conductance, which has now been
investigated in great detail
theoretically.\cite{Ini,Tanaka,Eschrig-rev} We shall however
concentrate on the spin currents in this paper. These surface
states can still generate the spin currents $J^z_y$ (and in
principle also $J^x_y$, but see below). However, as we shall see,
this is {\em not} the only contribution.  For the case of
$|\Delta_s|>|\Delta_p|$, the superconductor belongs to the trivial
category and it is expected that the surface bound states do not
survive. An immediate question is, at the transition point for
$|\Delta_p|=|\Delta_s|$, whether the spin current drops to zero
abruptly, signaling the transition of topology, or it is smoothly
decreasing towards zero, coinciding with the pure $s$-wave case
only when $\Delta_p$ vanishes.
In the following, we shall consider the above questions by
evaluating quantities such as the momentum- and spin-resolved
density of states and the spin current of the NcSC with varying
ratio $|\Delta_p|/|\Delta_s|$. We find that nonzero spin current
$J^z_y$ is dictated only by the broken symmetry and can be finite
for both the topological trivial and non-trivial superconductors.
The contribution to this spin current turns out to arise also from
non-topological {\em continuum} states with energies between the
two gaps $|\Delta_p \pm \Delta_s|$. These states are present in
both the topological trivial and non-trivial superconductors, and
they are not required by topology. (We recall here the analogous
situation that a finite spin Hall conductivity is possible from
the Kubo formula\cite{Murakami} yet the model can belong to the
topologically trivial $Z_2$ class\cite{TI3D}). Lastly, though in
principle $J^x_y$, and $J^y_x$ are allowed by symmetry, we found
that they vanish within our calculations.

\section{quasi-classical Green's function associated with NcSC}

Now we use the quasiclassical Green's functions to investigate the
surface of a clean NcSC. Here we shall employ the so-called
exploding and decaying tricks,\cite{Thuneberg,Fogelstrom,Yip1997}
which is related to the projector formalism initiated by
Shelankov.\cite{Shelankov}  This method is different from the
approach which employs the Riccati transformation.\cite{Eschrig09}
The Matsubara Green's function
$\hat{\mathrm{g}}(\hat{\rm{k}},\epsilon_n,\rm{R})$ in spin and
particle-hole space  satisfies the Eilenberger
equation,\cite{Serene}

\be
    [ i \epsilon_n \tau_3 - \hat \Delta, \hat \mathrm{g} ] + i \vec v_f
    \cdot \vec \nabla_{\rm{R}} \hat \mathrm{g} = 0\:, \label{qc1}
\ee with the normalization condition

\be
    \hat \mathrm{g}^2 = - \pi^2\hat{1}\:. \label{qc2}
\ee Here $\epsilon_n$ and $\hat{\rm{k}}$ denote the Matsubara
frequency and momentum direction associated with the quasiparticles,
respectively. $\rm{R}$ represents the spatial position. The set of
matrix $\{1,\vec{\tau}\}$ is used in the particle-hole sector while
$\{1,\vec{\sigma}\}$ serve for the usual spins. In this
representation, the 4$\times$4 pairing order parameter can be
written as

\be \hat \Delta = \left(
\begin{array}{cc} 0 & \underline{\Delta} \\ - \underline{\Delta}^\dag
& 0 \end{array} \right)\:, \label{OP} \ee where the 2$\times$2
matrix
$\underline{\Delta}$=$[\Delta_s+\Delta_p\vec{\rm{d}}(\vec{\rm{k}})\cdot\vec{\sigma}](i\sigma_y)$
in the usual spin representation. Here $\hat{\bullet}$ and
$\underline{\bullet}$ are to denote the 4$\times$4 and 2$\times$2
matrices, respectively. The triplet order parameter we consider is
of Rashba form $\vec{\rm{d}}=(-k_y,k_x)$. $\Delta_s$ and
$\Delta_p$ can be both taken real since we assumed TRS. Without
loss of generality, they will both be assumed positive. There are
two energy gaps, $|\Delta_p\pm\Delta_s|$, associated with the
quasiparticles in the superconducting states.\cite{2Gap} One
should also note that we have not included the spin-orbital
coupling term which may arise from the lack of inversion symmetry
in the normal state. Therefore, only one Fermi surface and its
associated Fermi velocity $v_f$ are needed. Qualitative effects of
including the spin-orbital coupling will be discussed later in the
paper.  For simplicity, we shall also ignore the spatial
variations of the order parameters $\Delta_{s,p}$.


We consider the geometry shown in Fig.\ \ref{fig1}. The incoming and
reflecting quasiparticles have the momentum $\underline{\rm{k}}$ and
$\rm{k}$, respectively. Here $k_x>0$ and $\underline{k}_x<0$. We
label positions along the quasiparticle path consisting of each
pairs of $\hat k$ and $\underline{\hat k}$ by $u$, with $u< 0$ ($u
> 0$) labels the part for $\hat k$ ($\underline{\hat k}$). The
quasiclassical Green's function at the surface
$\hat{\mathrm{g}}(0)$ can be obtained in terms of the decaying and
exploding solutions to eq.\ (\ref{qc1}) (c.f. Ref.\cite{Yip1997})

\be
    \hat{\mathrm{g}}(0)=-i\pi[\hat{A}(\rm{k}),\hat{B}(\underline{\rm{k}})]
    \{\hat{A}(\rm{k}),\hat{B}(\underline{\rm{k}})\}^{-1}\:,\label{gf1}
\ee where the matrix $\hat{A}=\hat{a}^{++}+\hat{a}^{--}$ and
$\hat{B}=\hat{b}^{++}+\hat{b}^{--}$. Below the 4 matrix solutions
$\hat{a}^{\pm\pm}$ and $\hat{b}^{\pm\pm}$ are explicitly shown,

\bea
    \hat{a}^{\pm\pm}=(\frac{1\pm\hat{d}\cdot\hat{\sigma}\tau_3}{2})\label{gf2}
    [-i|\Delta_{\pm}|^2\tau_3-i(\alpha_{\pm}+\epsilon_n)
    (\Delta_s+\Delta_p\hat{d}\cdot\vec{\sigma})\sigma_y\tau_{+}\\\nonumber
    +i(\alpha_{\pm}-\epsilon_n)\sigma_y
    (\Delta^{*}_s+\Delta^{*}_p\hat{d}\cdot\vec{\sigma})\tau_{-}]\:,\\
    \hat{b}^{\pm\pm}=(\frac{1\pm\hat{d}\cdot\hat{\sigma}\tau_3}{2})
    [i|\Delta_{\pm}|^2\tau_3-i(\alpha_{\pm}-\epsilon_n)
    (\Delta_s+\Delta_p\hat{d}\cdot\vec{\sigma})\sigma_y\tau_{+}\label{gf3}\\
    +i(\alpha_{\pm}+\epsilon_n)\sigma_y
    (\Delta^{*}_s+\Delta^{*}_p\hat{d}\cdot\vec{\sigma})\tau_{-}]\:,\nonumber
\eea where the notation we used for the three 4$\times$4 matrix
$\hat{\sigma} \equiv (\sigma_x,\sigma_y\tau_3,\sigma_z)$ should be
noticed. The parameters
$\alpha_{\pm}=\sqrt{\epsilon_n^2+|\Delta_{\pm}|^2}$, where
$\Delta_{\pm}=\Delta_s\pm\Delta_p$. More details can be found in
Appendix \ref{trick}.


\section{Surface bound states and spin currents}

Equipped with  Eq.\ (\ref{gf1}), (\ref{gf2}), and (\ref{gf3}), we
are ready to investigate the surface bound states and spin
currents at the surface. We do so by evaluating the spin and
momentum- resolved densities of states along different spin
projection directions.  For example, for a given momentum, the
density of states at the surface for spins along positive
(negative) $\rm{z}$ axis is given by
$\rho_{\rm{z\pm}}(\epsilon)=-\frac{N_F}{\pi}\rm{Im}\rm{Tr}[\frac{1\pm\sigma_{z}}{2}
\underline{\mathrm{g}}^R(\hat k,\epsilon;x=0)]$, where
$\underline{\mathrm{g}}^R =
\underline{\mathrm{g}}|_{i\epsilon_n\rightarrow\epsilon+i\delta,
\epsilon_n > 0}$ is the retarded Green's function. Similar formulas
apply by replacing $z$ with the other corresponding directions. Here
$\underline{\mathrm{g}}$ is the 2$\times$2 matrix in spin space
given by the upper left block of $\hat{\mathrm{g}}$ in the Nambu
space.

We shall show the results for representative cases of nontrivial
Z$_2$ with $|\Delta_p/\Delta_s| = 2$ and trivial Z$_2$ with
$|\Delta_p/\Delta_s|=1/2$. Fig.\ref{SpinZ} shows the spin-resolved
density of states $\rho_{\rm{z},\pm}$ with $\phi$=$\pm\frac{\pi}{6}$
and $|\Delta_p/\Delta_s|=2$. The momentum $k_y=k_F\sin\phi$ is a
good quantum number as a result of the translational invariance
along $\hat{\rm{y}}$. The peaks below the gap indicate the surface
bound state energy corresponding to

\be
    \rm{det}\left[\{\hat{A}(\rm{k}),
    \hat{B}(\underline{\rm{k}})\}\right]
    |_{i\epsilon_n\rightarrow\epsilon+i\delta}=0\:.\label{boundstateE}
\ee  In the case of pure p-wave ($\Delta_s = 0$), the bound state
spectrum is given by $E_{S_{\rm{z}}=\pm}(k_y)=\mp|\Delta_p|
\sin\phi$.\cite{Eschrig,Lu3} In this limit these surface
quasiparticles have a fixed spin orientation. Hence $\rho_{z
\pm}(\epsilon)$ consists of a single delta function peak at
$\epsilon = \mp |\Delta_p|\sin\phi$, whereas $\rho_{x,y
\pm}(\epsilon)$ are both identically zero. In contrast, in
Fig.\ref{SpinZ} (a) and (b), we find two subgap peaks, though with
large differences in height. These indicate that the
quasiparticles in the NcSC are no longer eigenstates of
$S_{\rm{z}}$ in the presence of order parameter mixing. These
minor peaks in $\rho_{z \pm}(\epsilon)$ vanish eventually as
$|\Delta_p|/|\Delta_s|$ is increased toward infinity. Moreover,
$\rho_{x,y \pm}(\epsilon)$ (Fig \ref{SpinX}, \ref{SpinY}) are also
finite at the bound state energies. This indicates that the
quantization axes for the spins are not aligned with $\hat z$.
Furthermore, the spin resolved density of states are finite also
for energies larger than $|\Delta_-|$.  This contribution is a
continuous (not delta) function in energy, indicating that this
contribution to the spin densities and spin currents arises from
{\em continuum} states not bound near the surface. We note that
kinks also appear in these density of states at the gap edges
where $\epsilon= \pm |\Delta_{\pm}|= \pm |\Delta_p\pm\Delta_s|$.
In this case, they are $\pm |\Delta_s|$ and $\pm 3|\Delta_s|$,
respectively.

For a given angle $\phi$ which parameterizes the quasiparticle path,
the surface bounds states appear in pairs of equal but opposite
energies. The bound state energy for the $E>0$ branch versus the
angle $\phi$ is plotted in Fig.\ref{Andreev}. It can be seen that
the Andreev bound states are pushed toward the band edge by the
$s$-wave pairing order parameter. We study in addition
(Fig.\ref{Andreev}, right panel)
the spin polarization $S_i \equiv
\frac{\rho_{i,+}-\rho_{i,-}}{\rho_{i,+}+\rho_{i,-}}$ of the Andreev
bound state versus the angle $\phi$. The sum in the denominator is
in fact independent of the direction $i$ and equals the density of
states. The $S_i$'s in some sense describe the spin direction of the
Andreev bounds state.\cite{footnote} With increasing
$\Delta_p/\Delta_s$, $S_z$ approaches a step function while $S_x$
approaches zero. Thus we can conclude that the effect of $s$-wave
order parameter is to tilt the spin of Andreev bound states toward
the x axis.

The spin current density $J^z_y$ at the surface is obtained via the
expression,

\be
  J^z_y (x=0)=\frac{\hbar}{2}N_F v_F\int\frac{d\phi}{2\pi} (\sin\phi)
  T\sum_n{\rm{Tr}}
  \left[
  {\sigma}^z\underline{g}(\hat k, \epsilon_n;0)
  \right]\:,
  \label{spincurrent}
  \ee
The summation over the Matusbara frequency ($T \sum_n \rm{Tr} [ ..
\underline{g}]$) can be replaced by the integral \bdm \int
\frac{d\epsilon}{2\pi} \rm{Im} \rm{Tr} [ ..
\underline{g}^R(\epsilon)]\tanh(\frac{\epsilon}{2T}) \edm with
respect to the real frequency $\epsilon$. We shall first note an
elementary symmetry relation followed from TRS of our
superconducting state

\be
    \rho_{\hat{\rm{n}},\pm}(k_y,\epsilon)=
    \rho_{\hat{\rm{n}},\mp}(-k_y,\epsilon)\:,\label{TRS_rho}
\ee which is  valid for all spin projections $\hat{\rm{n}}$. This
can be seen in Fig.\ref{SpinZ}, Fig.\ref{SpinX} etc. From Fig
\ref{SpinZ}, we see that we also have, in addition, the symmetry
$\rho_{\rm{z}\pm}(k_y,\epsilon)=\rho_{\rm{z}\pm}(-k_y,-\epsilon)$
(and consequently
$\rho_{\rm{z}+}(k_y,\epsilon)=\rho_{\rm{z}-}(k_y,-\epsilon)$). The
asymmetry between $\rho_{\rm{z}+}(k_y,\epsilon)$ and
$\rho_{\rm{z}-}(k_y,\epsilon)$ (and between
$\rho_{\rm{z}+}(k_y,\epsilon)$ and
$\rho_{\rm{z}+}(-k_y,\epsilon)$) causes a finite spin current
$J^z_y$ in the topological nontrivial cases as well as in the
trivial cases, whereas eq (\ref{TRS_rho}) guarantees that all spin
accumulations are zero. For spin projections along $\hat{\rm{x}}$
and $\hat{\rm y}$ as in Fig.\ref{SpinX} and Fig.\ref{SpinY}, we
have in addition
$\rho_{\rm{x,y}+}(k_y,\epsilon)=\rho_{\rm{x,y}+}(k_y,-\epsilon)$,
(and similarly for $+ \to -$) which forbids both the spin currents
$J^x_y$ and $J^y_x$.


In the trivial Z$_2$ case where we choose
$|\Delta_p|$=$|\Delta_s|$/2, no subgap state is found as in
Fig.\ref{Trivial}(a). However, there is still an asymmetry between
$\rho_{\rm{z}+}(k_y,\epsilon)$ and $\rho_{\rm{z}-}(k_y,\epsilon)$
for the continuum states between the two gaps, so that these
states continue to contribute to $J^z_y$. The quasiparticles upon
reflection at the surface is now analogous to transmission through
a Josephson junction between two unequal-gap SC's with a relative
phase difference,\cite{unequal} where inter-gap continuum states
contribute to a finite current. We suggest here that the
contribution to the spin current can be pictured in similar
manner. With  Eq.\ (\ref{spincurrent}), we find that $J^z_y$ is
indeed nonvanishing. This is justified in the real frequency
domain as shown in Fig.\ref{Trivial}(b) where a finite
contribution to $J^z_y$ is found at energies $\epsilon$ between
the gaps $ - |\Delta_+| < \epsilon < -|\Delta_-|$ ($  |\Delta_-| <
\epsilon < |\Delta_+|$).  Note that there is no contribution from
states with $|\epsilon| > |\Delta_+|$ (These two statements also
hold for the topologically non-trivial case).

In Fig.\ref{Jzy}, the values of $J^z_y(x=0)$ as a function of the
triplet to singlet order parameter ratio $|\Delta_p|/|\Delta_s|$
is presented. $J^z_y(x=0)$ increases from zero in pure $s$-wave
case toward the value obtained in the pure triplet case. In the
topologically trivial regime, $|\Delta_p|<|\Delta_s|$,
$J^z_y(x=0)$ seems to be quadric in $|\Delta_p|$; while in the
topologically nontrivial regime, it is roughly linearly with
$|\Delta_p|$. The asymptotic value for $J^z_y(x=0)$ coincides with
our previous results.\cite{Lu3} Fig.\ref{Izy} shows the values of
the total spin current $I^z_y \equiv \int_0^{\infty} dx J^z_y(x)$.

\section{discussions and conclusions}

The pairing Hamiltonian can be conveniently written in the
helicity basis,\cite{YipPRB2002}
$H_{\Delta}=\frac{1}{2}\sum_{\bf{k}}(i\Delta_{+}e^{-i\phi_{\rm{k}}}a^{\dag}_{\rm{k}+}a^{\dag}_{-\rm{k}+}-
i\Delta_{-}e^{i\phi_{\rm{k}}}a^{\dag}_{\rm{k}-}a^{\dag}_{-\rm{k}-})
\  + h.c.$. Here $\Delta_{\pm}$ are both real numbers so that the
above is time-reversal invariant. $\phi_{\rm{k}}$ is the angle
between the momentum $\rm{k}$ and the x axis. The relation
$\Delta_{s,p}=(\Delta_{+}\pm\Delta_{-})/2$ can be obtained upon
transforming back to the normal spin basis
$\{a^{\dag}_{\rm{k}\uparrow},a^{\dag}_{\rm{k}\downarrow}\}$. Thus
the pure $p$-wave Rashba triplet SC and the pure $s$-wave singlet
SC are recovered when $\Delta_{+}=\mp\Delta_{-}$, respectively;
while the case with $|\Delta_{+}|\neq|\Delta_{-}|$ belongs to the
NcSC. The nontrivial Z$_2$ class corresponds to
$\Delta_{+}\Delta_{-}<0$.\cite{YipTopology,ZhangTopology}

The pure s and p-wave cases, $|\Delta_{+}|=|\Delta_{-}|$, are well
understood from the previous papers.\cite{Eschrig,Lu3} The main
difference between the pure triplet case and the present NcSC is
that the surface Andreev bound states do not have a fixed spin
projection. Moreover, besides the surface bound states, the
continuum states between the two gaps, $|\Delta_{\pm}|$, also
contribute to the spin current and are solely responsible for its
nonvanishing value in the topologically trivial case.

In our calculations, where the normal state spin splitting is not
included, only the component $J^z_y$ is nonvanishing  while the
other components do not appear, even though the quasiparticles
have their spins not aligned with the $\hat z$ axis. Mixed-parity
SC order parameter along with the presence of Rashba spin-orbital
interaction accounting for the absence of inversion symmetry may
be closer to the physically accessible regime.\cite{Sergienko}
However, the inclusion can in principle generate nonzero spin
currents even in the normal state under equilibrium.\cite{Rashba}
Even so, the spin current obtained in the normal state for a
spin-split band is \emph{much smaller} than what we have obtained
in our mixed-parity superconducting state without the
corresponding spin-orbital coupling, provided that the factor
$T_c{E_F^2}/\alpha^3$ is sufficiently large,\cite{Rashba,Eschrig}
which is presumably true in usual situation when the spin
splitting energy $\alpha\ll{E_F}$. Moreover, inclusion of the
spin-orbital coupling in the normal state will make the spin
current ill-defined because of the lack of spin conservation. In
Ref.\cite{Eschrig}, the authors define the \textit{real} spin
current in superconductors by subtracting the contribution
obtained in the normal state. They then obtained also finite
$J^x_y$ and $J^y_x$.

In conclusion, we use the exploding-decay tricks to obtain the
quasi-classical Green's functions associated with a
singlet-triplet mixed noncentrosymmetric superconductor. For the
topologically nontrivial NcSC, we obtain a pair of Andreev bound
states without a fixed spin projection and a consequent spin
current $J^z_y$. For the topologically trivial NcSC, a finite spin
current $J^z_y$ remains even though no Andreev bound state can be
found.

\begin{center}
{\bf Acknowledgment}
\end{center}

This work is supported by the National Science Council of Taiwan.
C.K.L. is also supported by the overseas postdoctoral fellowship
program from NSC Taiwan.

\appendix

\section{exploding \& decaying trick in NcSC}\label{trick}

\begin{figure}
\input{epsf}
\includegraphics[scale=0.4]{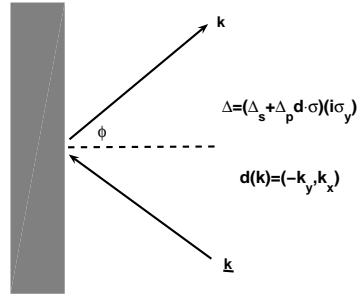}
\caption{\label{fig1} The interface between vacuum(gray) and a
noncentrosymmetric superconductor(white) with singlet-triplet
mixed order parameter specified by text. The directions are
defined by the shown axis. The quasicparticle is incident from the
lower right of xy plane along the path denoted by
$\hat{\underline{k}}$. $\phi$ is the reflection angle between the
x-axis and the outgoing path $\hat{\rm{k}}$.}
\end{figure}

\begin{figure}
\input{epsf}
\includegraphics[scale=0.4]{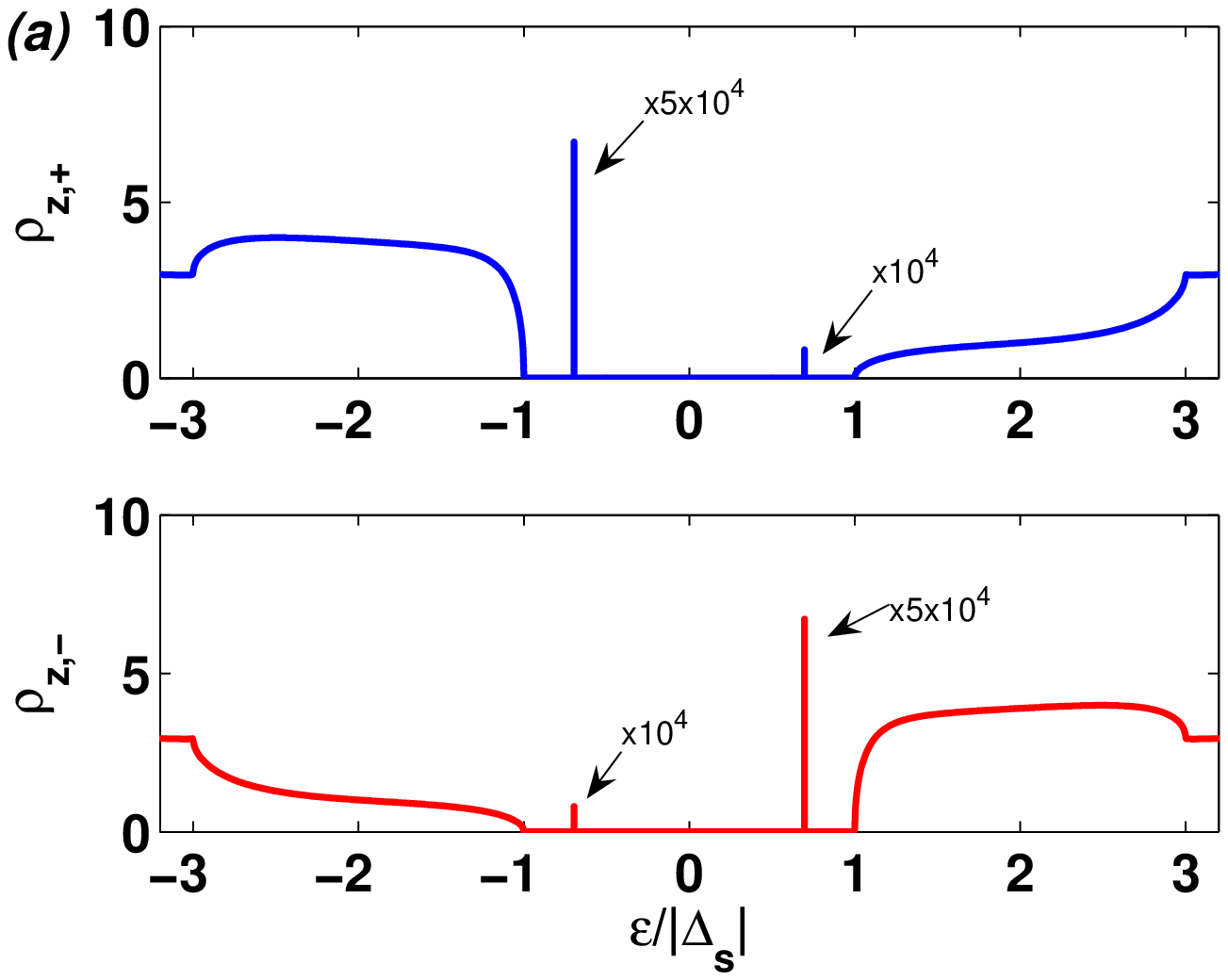}
\includegraphics[scale=0.4]{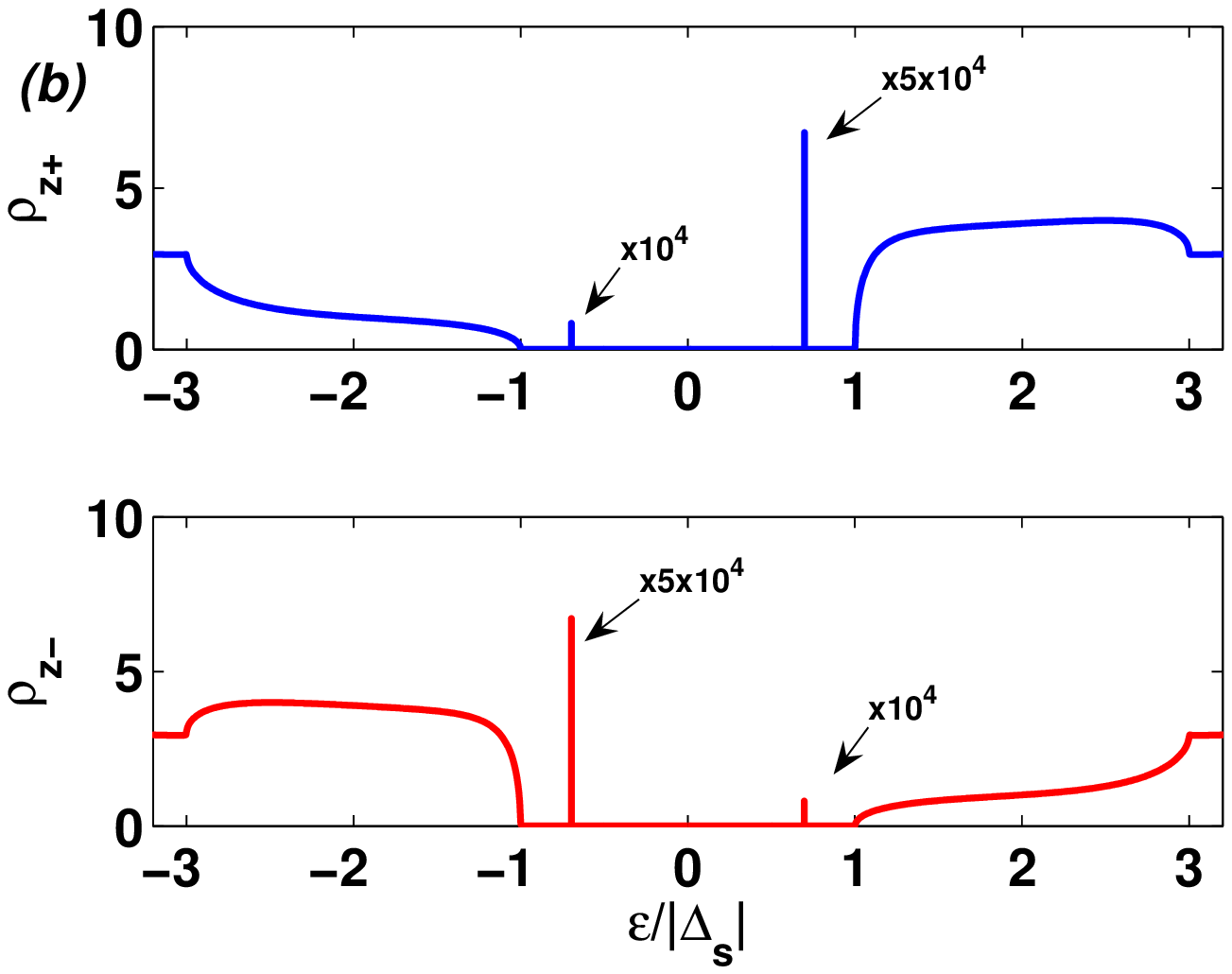}
\caption{\label{SpinZ} (Color online) Momentum and spin-resolved
surface densities of states  $\rho_{\rm{z},\pm}$ in unit of
$\frac{N_f}{\pi}$ with $|\Delta_p|$=2$|\Delta_s|$.
(a)$\phi=\frac{\pi}{6}$. (b)$\phi=-\frac{\pi}{6}$. Note that the
upper(lower) plot in (a) is identical to the lower(upper) one in (b)
due to the time-reversal symmetry. The numerical values associated
with the subgap peaks are related to the small imaginary number
$\delta=10^{-6}$ we used in the transformation
$i\epsilon_n\rightarrow\epsilon+i\delta$, which is also true in
Fig.\ref{SpinX}.}
\end{figure}

\begin{figure}
\input{epsf}
\includegraphics[scale=0.4]{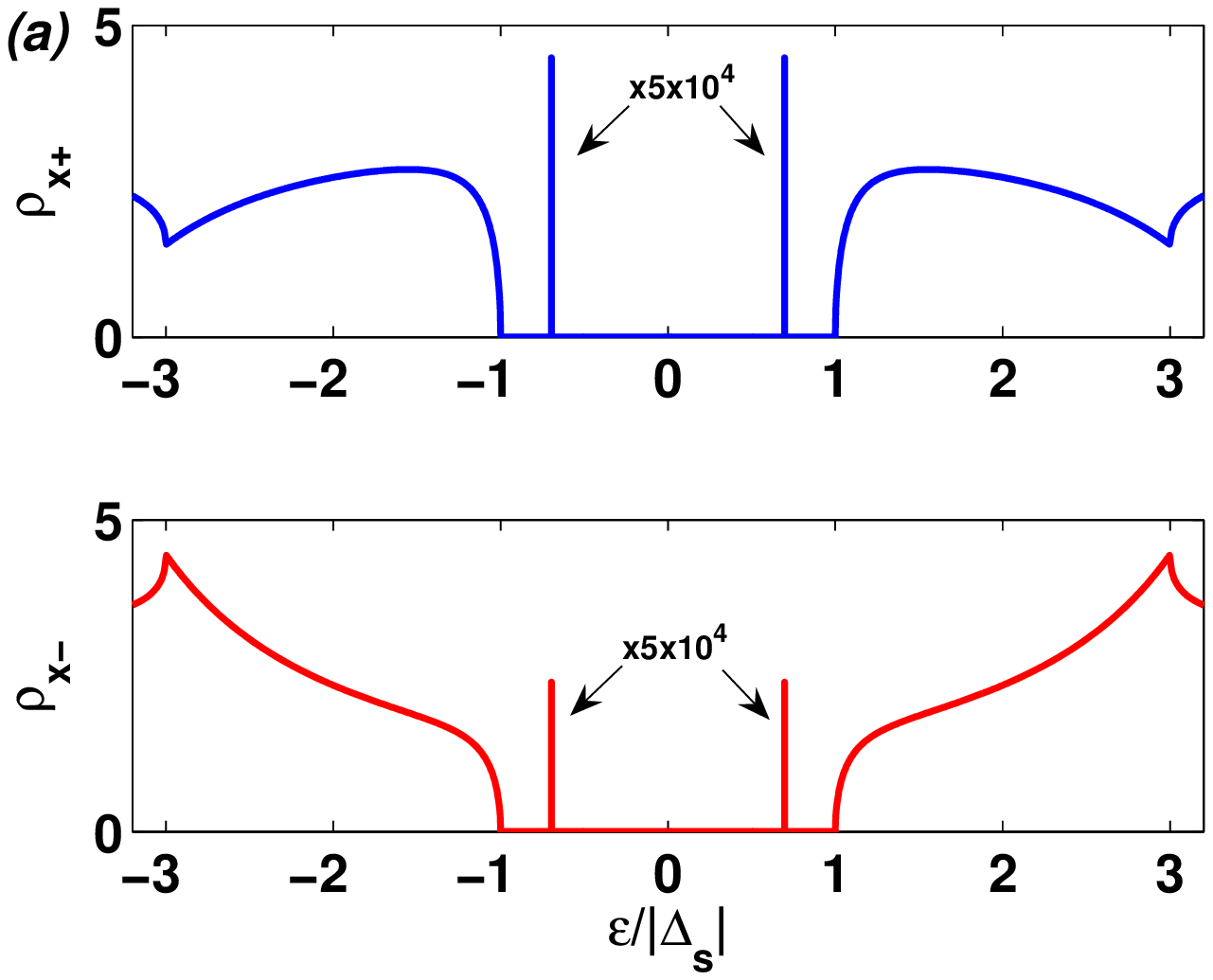}
\includegraphics[scale=0.4]{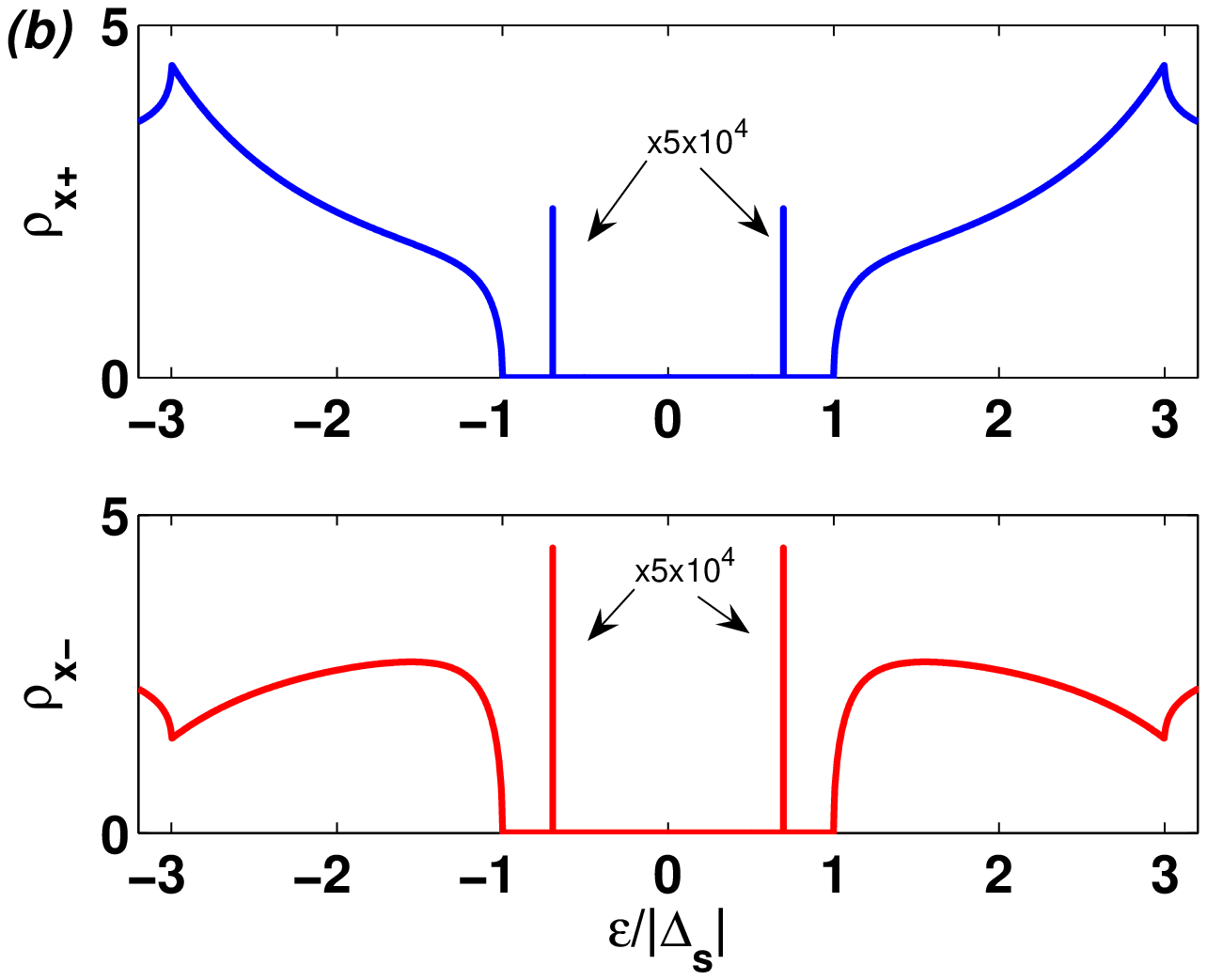}
\caption{\label{SpinX} (Color online) Momentum and spin-resolved
surface densities of states $\rho_{\rm{x},\pm}$ in unit of
$\frac{N_f}{\pi}$ with $|\Delta_p|$=2$|\Delta_s|$. (a)
$\phi=\frac{\pi}{6}$. (b) $\phi=-\frac{\pi}{6}$. Besides the same
symmetry followed by TRS, it shows in additional symmetry with
$\epsilon\leftrightarrow{-}\epsilon$ which forbids the spin current
$J^x_y$.}
\end{figure}

\begin{figure}
\input{epsf}
\includegraphics[scale=0.4]{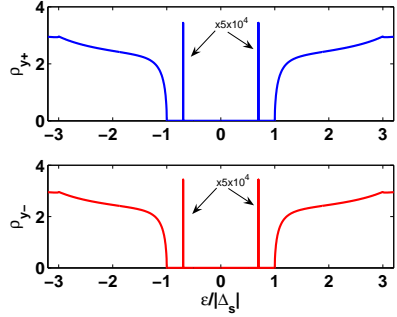}
\caption{\label{SpinY} (Color online) Same as Fig \ref{SpinX}
except here that $\rho_{y,\pm}$ is shown. The results for $\phi =
\pm \frac{\pi}{6}$ are identical.}
\end{figure}

\begin{figure}
\input{epsf}
\includegraphics[scale=0.4]{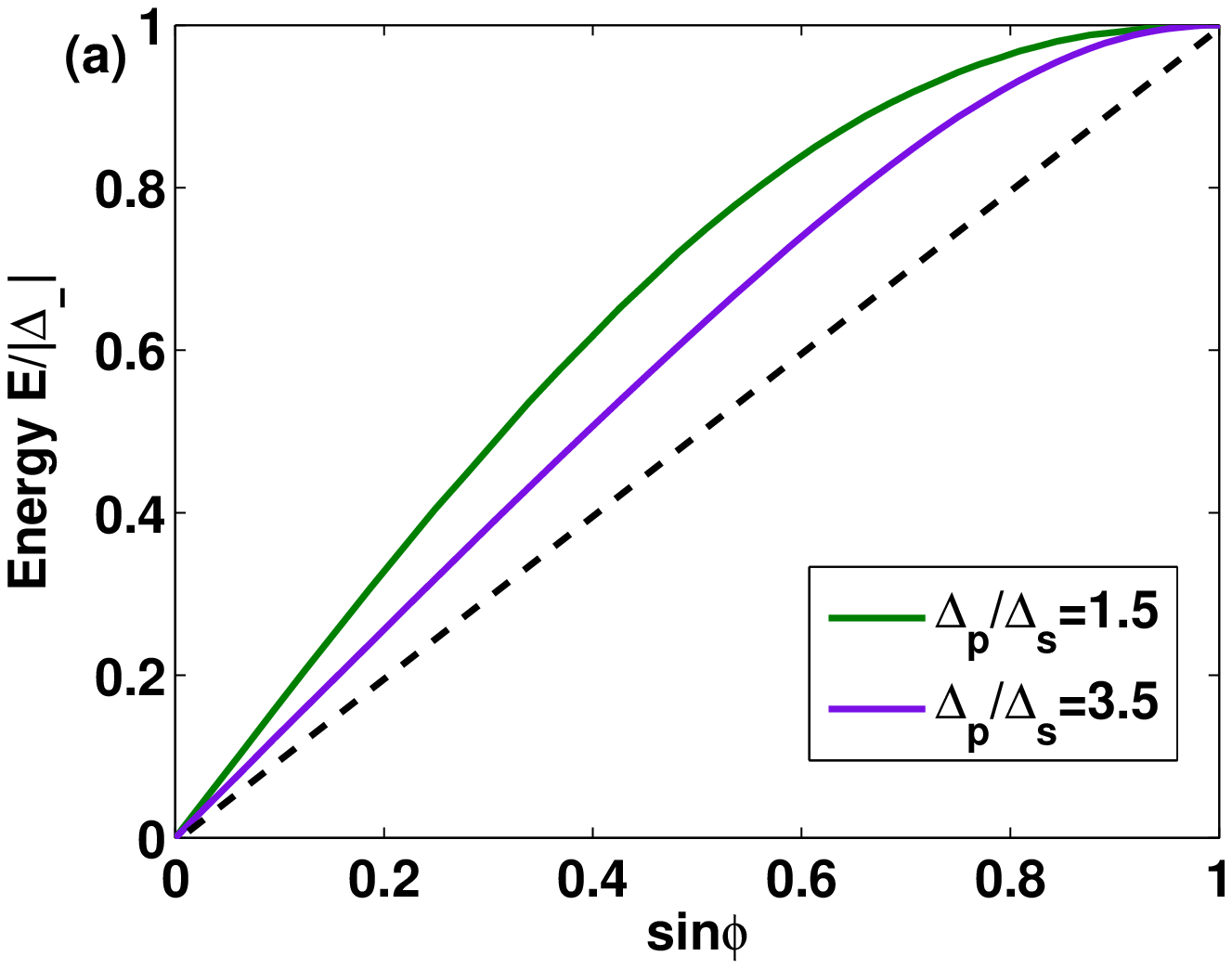}
\includegraphics[scale=0.4]{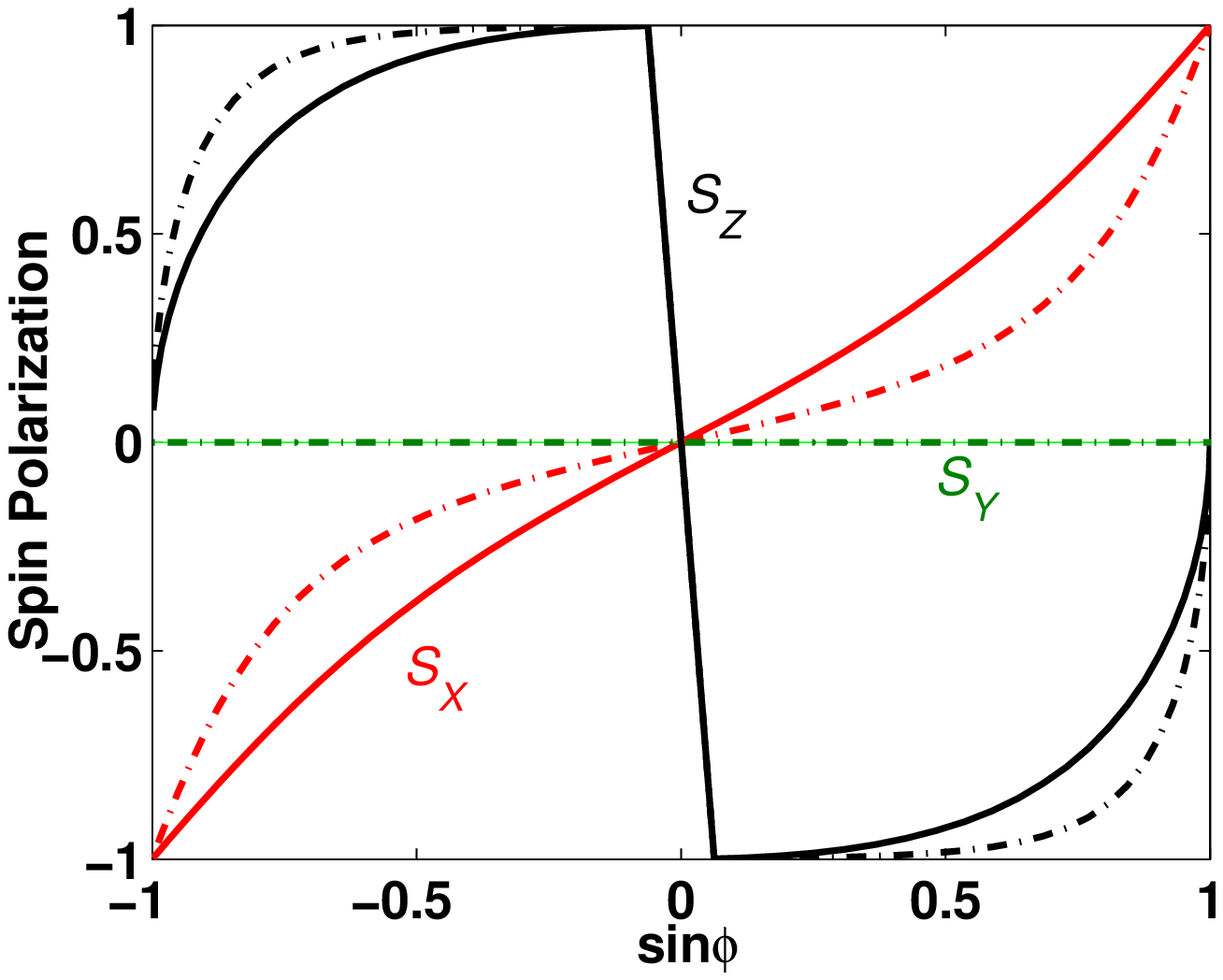}
\caption{\label{Andreev} (Color online) Left: Positive energy branch
of the surface Andreev bound states, corresponding to the zeros of
the anticommutators in Eq.\ (\ref{boundstateE}), as a function of
the angle $\phi$ for two ratios of $\Delta_p/\Delta_s$. The
upper(green) and lower(purple) are for the ratios of 1.5 and 3.5,
respectively. $E$ is measured in terms of $\Delta_{-}$. As the
triplet component becomes larger, the bound state behaves more like
in the pure triplet case where $E=|\Delta_p|\sin\phi$ (the black
dashed line). Right: Spin polarization of Andreev bound states
(defined in text) versus the angle $\phi$. Solid lines are for the
ratio $\Delta_p/\Delta_s$ of 1.5 and the dashed ones are for 3.5.
The direction of spin polarization evolves in the xz plane as $\phi$
varies. For smaller angles or larger $\Delta_p/\Delta_s$, the bound
states are more spin-polarized along z as in the pure triplet case.}
\end{figure}

\begin{figure}
\input{epsf}
\includegraphics[scale=0.4]{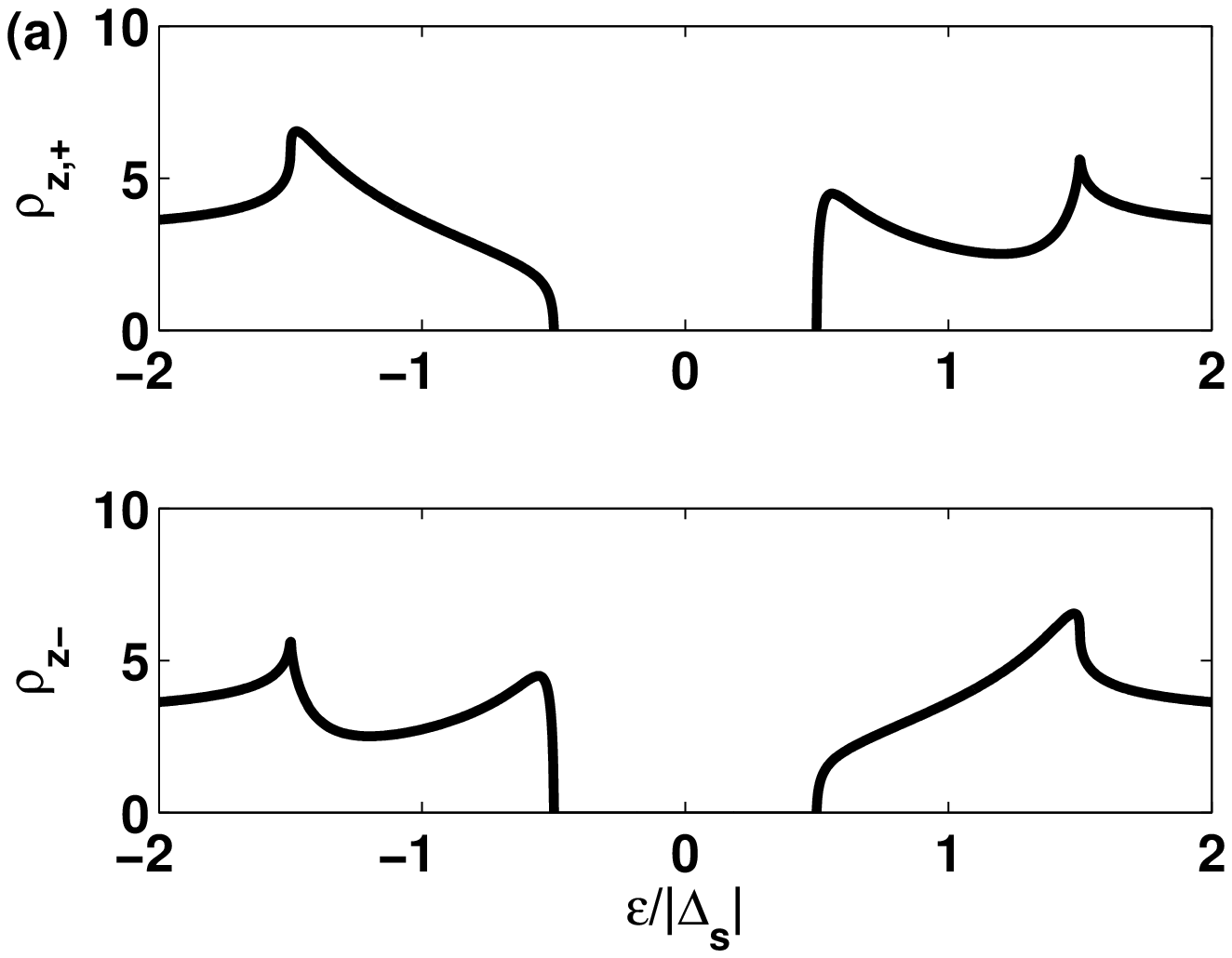}
\includegraphics[scale=0.4]{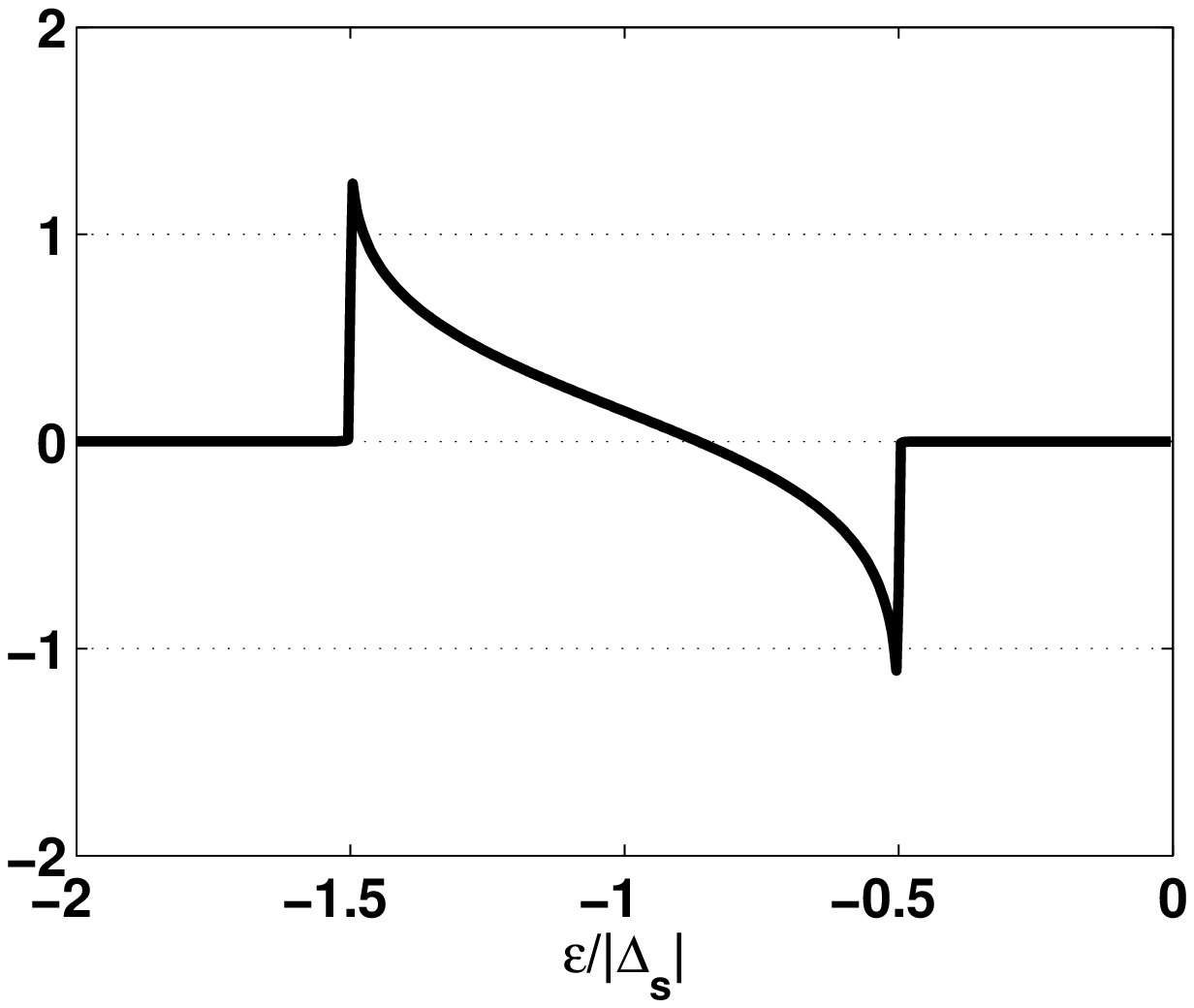}
\caption{\label{Trivial} Topologically trivial NcSC with
$|\Delta_p|$=0.5$|\Delta_s|$. (a): Spin-resolved surface density of
states $\rho_{z\pm}$ associated with $k_y=k_F/2$. No subgap states
appear. (b): Non-vanishing inter-gap direction-integrated
contributions
$\int_{0}^{\pi/2}d\phi(\sin\phi){\rm{Tr}}\{\sigma_z\tau_3[-\frac{1}{\pi}{\rm{Im}}\
{\hat{\rm{g}}^R}(\epsilon,\phi)]\}$ to $J^z_y$ in the unit of
$\hbar{v_FN_F}|\Delta_s|$. }
\end{figure}

\begin{figure}
\input{epsf}
\includegraphics[scale=0.4]{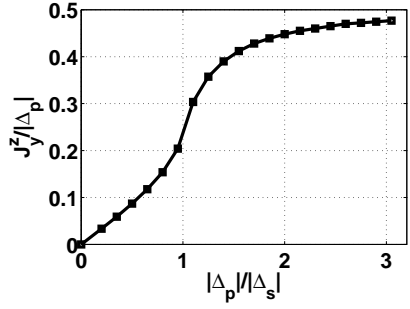}
\caption{\label{Jzy} Surface spin current density $J^z_y$ (in unit
of $\hbar{v_FN_F}|\Delta_p|$) versus the order parameter ratio
$|\Delta_p|/|\Delta_s|$ at a temperature $T=|\Delta_s|/100$. In
the large $|\Delta_p|$ limit,
$J^z_y=\frac{1}{2}\hbar{v_FN_F}|\Delta_p|$, while the spin current
vanishes in the pure singlet case. As $|\Delta_p|$ increases from
the topologically trivial to nontrivial regimes, $J^z_y$ follows
different dependence on $|\Delta_p|$ around the transition
$|\Delta_p|=|\Delta_s|$.}
\end{figure}

\begin{figure}
\input{epsf}
\includegraphics[scale=0.4]{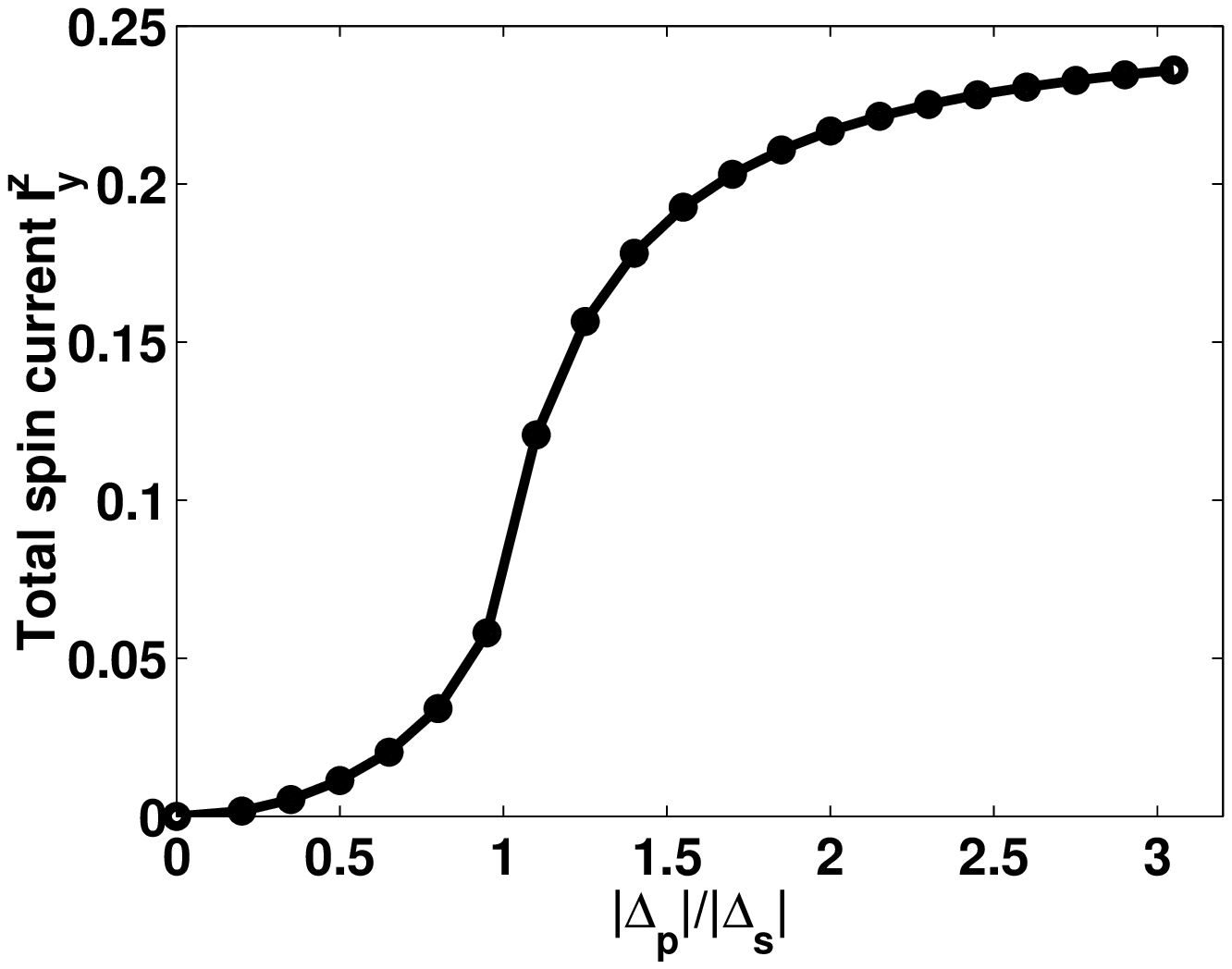}
\caption{\label{Izy} Total surface spin current density $I^z_y$
(in the unit of $\hbar^2 {N_F v_F^2 }$) versus the order parameter
ratio $|\Delta_p|/|\Delta_s|$ at a temperature
$T=|\Delta_s|/100$.}
\end{figure}

Here we shall present our scheme for obtaining the solutions given
in Eq.\ (\ref{gf1}) to the Eilenberger equations Eq.\ (\ref{qc1})
and (\ref{qc2}) for the NcSC in the presence of a boundary as
shown in Fig.\ref{fig1}. Eq.\ (\ref{gf1}) is expressed in terms of
the exploding and decaying solutions Eq.\ (\ref{gf2}) and
(\ref{gf3}) of an auxiliary problem that is identical to the
present NcSC for $x>0$. To make our reasoning clearer, we first
consider the special case where $\hat{d}$ coincides with
$\hat{\rm{z}}$, then eventually obtain the Green's function for
general $\hat d$. (A similar method to obtain the quasiclassical
Green's function for a uniform non-centrosymmetric superconductor
has been also used in Ref.\cite{QcGF}.  Our treatment extends to
the non-uniform case and we also point out  some useful
mathematical relations not noted in there). The matrix
$\hat{\rm{M}}=(i\epsilon_n-\hat{\Delta})$ appearing in the
commutator in Eq.\ (\ref{qc1}) becomes, after an obvious
rearrangement of rows and columns, block-diagonalized. Explicitly,
with $u=\hat{\rm{k}}\cdot\rm{R}$, it has the form

\bea
    [ \left(\ba{cc} \underline{\rm{M}}^{++} & 0\\
    0 & \underline{\rm{M}}^{--}\ea \right),
    \left(\ba{cc} \underline{\rm{g}}^{++} &\underline{\rm{g}}^{+-}\\
    \underline{\rm{g}}^{+-} & \underline{\rm{g}}^{--}\ea \right)]
    +iv_F\partial_u
    \left(\ba{cc} \underline{\rm{g}}^{++} &\underline{\rm{g}}^{+-}\\
    \underline{\rm{g}}^{+-} & \underline{\rm{g}}^{--}\ea \right)
    =0\:,
    \label{block}
\eea where the diagonal elements of $\hat{\rm{M}}$ are

\be
    \underline{\rm{M}}^{\pm\pm} \equiv
    \left(\ba{cc}{i\epsilon_n}&\mp\Delta_{\pm}\\
    \pm\Delta^{*}_{\pm}&{-i\epsilon_n}\ea\right)\:.
    \label{defM}
\ee The parameters $\Delta_{\pm}=\Delta_s\pm\Delta_p$. By imposing
the spatial dependence of $e^{-\frac{2{\lambda}u}{v_F}}$ to
$\hat{g}$, Eq.\ (\ref{block}) becomes a  problem for finding
eigenvalues $\lambda$ and the corresponding eigenvectors. For
example the upper diagonal block ($++$) simplifies to
$[\underline{\rm{M}}^{++},\underline{\rm{g}}^{++}]-2i\lambda\underline{\rm{g}}^{++}=0$.
This relation is analogous to the pure s or p wave case and can be
solved in the same manner.  The eigenvalues $\lambda$ form the set
$\{0,0,\alpha_{+},-\alpha_{+}\}$ with $\alpha_{+} \equiv
\sqrt{\epsilon_n^2+|\Delta_{+}|^2}$ and the associated
eigenvectors are respectively the set of 2$\times$2 matrices
$\{-i\pi{\bf{1}},\underline{g}^{++}_b,\underline{a}^{++}_{\alpha_{+}},\underline{b}^{++}_{-\alpha_{+}}\}$.
The previous two are the "constant solutions" while
$\underline{a}^{++}_{\alpha_{+}}$ and
$\underline{b}^{++}_{-\alpha_{+}}$ are the decaying and exploding
ones, given by

\bea
    \underline{a}^{++}_{\alpha_{+}}=
    \left(\ba{cc}{-i|\Delta_{+}|^2}&-\Delta_{+}(\alpha_{+}+\epsilon_n)\\
    -\Delta^{*}_{+}(\alpha_{+}-\epsilon_n)&{i|\Delta_{+}|^2}\ea\right)\:,\label{g_a}\\
    \underline{b}^{++}_{\alpha_{+}}=
    \left(\ba{cc}{i|\Delta_{+}|^2}&-\Delta_{+}(\alpha_{+}-\epsilon_n)\\
    -\Delta^{*}_{+}(\alpha_{+}+\epsilon_n)&{-i|\Delta_{+}|^2}\ea\right)\:,\label{g_b}
\eea respectively. Besides, the following equality holds as well
as in e.g. Ref.\cite{Lu3}

\bea
    \underline{g}^{++}_b &=& -i\pi[\underline{a}^{++},\underline{b}^{++}]
    \{\underline{a}^{++},\underline{b}^{++}\}^{-1} \label{g_bulk} \\
    &=&\frac{-\pi}{\alpha_{+}}\underline{\rm{M}}^{++}\:
\eea where the subscripts for eigenvalues are omitted if no
confusion is generated. Similar results can be obtained in the
($--$) block by defining
$\alpha_{-}=\sqrt{\epsilon_n^2+|\Delta_{-}|^2}$. As for the
off-diagonal blocks ($+-$) and ($-+$), they share the same set of
eigenvalues, $\{\pm\alpha_s,\pm\alpha_d\}$, with the subscripts
standing for the \textit{sum} and \textit{difference} by
$\alpha_{s,d}=\frac{\alpha_{+}\pm\alpha_{-}}{2}$. The eigenvectors
belonging to the block ($ij$) $=(+-)$ or $(-+)$ will be labelled
as
$\{\underline{a}^{ij}_{\alpha_{s,d}},\underline{b}^{ij}_{\alpha_{s,d}}\}$.
It should be reminded that the decaying (exploding) solutions are
associated with positive (negative) eigenvalues.

Since the product of two solutions to Eq.\ (\ref{block}) also
solves that equation, this product must be proportional to another
solution, or it must vanish, depending on whether the sum of the
corresponding eigenvalues coincides or not with any of the allowed
eigenvalues. More precisely, if $\underline{g}_{\lambda}^{ij}$ is
a solution in the $(ij)$ block with eigenvalue $\lambda$, then the
product
$\underline{g}_{\lambda_1}^{ij}\underline{g}_{\lambda_2}^{jk}$ is
identical to $\underline{g}_{\lambda_1+\lambda_2}^{ik}$ (up to a
proportionality constant) or zero depending on whether
$\lambda_1+\lambda_2$ coincides with one of the eigenvalues
associated with the block ($ik$). Using Eq.\ (\ref{g_bulk}), the
following relations can be easily shown,

\bea
    \underline{a}^{\pm\pm}\underline{g}^{\pm\pm}_b&=&i\pi\underline{a}^{\pm\pm}\:,\label{ag}\\
    \underline{b}^{\pm\pm}\underline{g}^{\pm\pm}_b&=&-i\pi\underline{b}^{\pm\pm}\:,\label{bg}\\
    \underline{a}^{+-}_{\alpha_{d,s}}\underline{g}^{--}_b&=&{\mp}i\pi\underline{a}^{+-}_{\alpha_{d,s}}\:,\label{apmg}\\
    \underline{a}^{-+}_{\alpha_{d,s}}\underline{g}^{++}_b&=&i\pi\underline{a}^{-+}_{\alpha_{d,s}}\:,\label{ampg}\\
    \underline{b}^{+-}_{\alpha_{d,s}}\underline{g}^{--}_b&=&\pm{i}\pi\underline{b}^{+-}_{\alpha_{d,s}}\:,\label{bpmg}\\
    \underline{b}^{-+}_{\alpha_{d,s}}\underline{g}^{++}_b&=&-{i}\pi\underline{b}^{-+}_{\alpha_{d,s}}\:,\label{bmpg}
\eea

Next we return to general $\vec{d}$ direction. Observing that
$\underline{\rm{g}}^{\pm \pm}_b$ in Eq.\ (\ref{g_bulk}) can in fact
be written as $[-\frac{\pi}{\alpha_{\pm}}\frac{1 \pm
\sigma_z\tau_3}{2}\hat{\rm{M}}(\hat{d}=\hat{\rm{z}})]$ in the
4$\times$4 Nambu notation, we see that the information about
$\vec{d}$ is embedded in the projection operator
$\rm{P}_{\pm}=\frac{1\pm\hat{d}\cdot\hat{\sigma}\tau_3}{2}$, where
$\hat{\sigma} \equiv \{\sigma_x,\sigma_y\tau_3,\sigma_z\}$,
\cite{Lu3} and in the order parameter $\hat \Delta$ which appears
inside $\hat M$.  From this, we obtain eqs (\ref{gf1}-\ref{gf3}).
Indeed, it is useful to note that $\rm{P}_{\pm}$ commute with
$\hat{\rm{M}}=(i\epsilon_n-\hat{\Delta_s}-\hat{\Delta_p})$.  It is
easy to see that $\rm{P}_{\pm}\hat{\rm{M}}$ are the homogeneous
solutions to eq (\ref{qc1}). We can also see that the decaying and
exploding solutions in Eq.\ (\ref{gf2}) and (\ref{gf3}) do satisfy
eq (\ref{qc1}), and one can further convince himself/herself by
checking the relations,
$(\hat{a}^{++})^2=(\hat{b}^{++})^2=(\hat{a}^{--})^2=(\hat{b}^{--})^2=0$,
and
$(\hat{\rm{g}}_b)^2=(\hat{\rm{g}}^{++}_b+\hat{\rm{g}}^{--}_b)^2=-\pi^2$,
again hold. By similar reasoning as the $\hat d = \hat z$ case, Eq.\
(\ref{ag})-(\ref{bmpg}) in the 4$\times$4 form still hold for
general $\vec{d}$.

Now we show how the present problem in Fig.\ref{fig1} can be solved
in terms of these auxiliary solutions. For the outgoing path denoted
by $\rm{k}$, the most general solution with the correct limit as $x
\to \infty$ is

\be
    \hat{\rm{g}}(u)=\hat{\rm{g}}_b(\rm{k})+c_1^{++}(u)\hat{a}^{++}(\rm{k})+c_1^{--}(u)\hat{a}^{--}(\rm{k})
    +c_1^{+-}(u)\hat{a}^{+-}_{\alpha_s}(\rm{k})+c_1^{-+}(u)\hat{a}^{-+}_{\alpha_s}(\rm{k})\:,\label{eq_out}
\ee with $u=\rm{R}\cdot{\hat{\rm{k}}}$ positive. For simplicity,
the spatial dependence $e^{-2\alpha_{+}u/v_F}$ associated with
$\hat{a}^{++}$ has been absorbed in the c-number coefficient
$c_1^{++}$, and similarly for the others. For the incoming path,

\be
    \hat{\rm{g}}(u)=\hat{\rm{g}}_b(\rm{\underline{k}})+c_2^{++}(u)\hat{b}^{++}(\rm{\underline{k}})
      +c_2^{--}(u)\hat{b}^{--}(\rm{\underline{k}})
    +c_2^{+-}(u)\hat{b}^{+-}_{\alpha_s}(\rm{\underline{k}})+
    c_2^{-+}(u)\hat{b}^{-+}_{\alpha_s}(\rm{\underline{k}})\:,\label{eq_in}
\ee with $u=\rm{R}\cdot{\hat{\rm{\underline{k}}}}$ negative here.
We should remark that in the above equations decaying/exploding
solutions associated with eigenvalue $\alpha_d$ in the
off-diagonal blocks has been excluded. We can see this by two
different arguments.  First, we can imagine that, far away from
the surface, the inversion symmetry is restored.  In that case
$\alpha_d \to 0$ and hence they would become constant solutions,
which cannot exist at $x \to \infty$.  Second, using relations
eq(\ref{ag}-\ref{bmpg}), we can check that their appearance would
violate the normalization condition (\ref{qc2}). Now putting $u=0$
in Eq.\ (\ref{eq_out}) and multiplying this equation with
$\hat{A}(\rm
k)=p_1\hat{a}^{++}+p_2\hat{a}^{--}+p_3\hat{a}^{+-}_{\alpha_s}+p_4\hat{a}^{-+}_{\alpha_s}$,
and using Eq.\ (\ref{ag}), (\ref{apmg}), and (\ref{ampg}), yields,

\be
    \hat{A}(\rm k)\hat{\rm{g}}(0)=i\pi\hat{A}(\rm k)\:,
\ee where the c-numbers $\{p_i;i=1,2,3,4\}$ are arbitrary.
Similarly, multiplying
$\hat{B}(\rm{\underline{k}})=q_1\hat{b}^{++}+q_2\hat{b}^{--}+
q_3\hat{b}^{+-}_{\alpha_s}+q_4\hat{b}^{-+}_{\alpha_s}$ with Eq.\
(\ref{eq_in}) and using Eq.\ (\ref{bg}), (\ref{bpmg}), and
(\ref{bmpg}), yields

\be
    \hat{B}(\rm{\underline{k}})\hat{\rm{g}}(0)=-i\pi\hat{B}(\rm{\underline{k}})\:,\\
\ee where again $\{q_i;i=1,2,3,4\}$ are arbitrary. Multiplying the
above two equations by $B(\rm{\underline{k}})$ and $A(\rm k)$
respectively and add,  one obtain the final expression in Eq.\
(\ref{gf1}). The choices for $p_i$ and $q_i$ are not restricted,
as long as the anticommutator $\{A(\rm k),B(\rm{\underline{k}})\}$
has a non-vanishing determinant and hence it is invertible. The
determinant vanishes when the energy coincides with the Andreev
bound state for a given $\phi$, but the invertibility can still be
guaranteed by adding a small imaginary number to the energy. Given
two sets of matrix, $\hat{\rm{A}}_1(\rm{k})$ and
$\hat{\rm{B}}_1(\rm{\underline{k}})$, $\hat{\rm{A}}_2(\rm{k})$ and
$\hat{\rm{B}}_2(\rm{\underline{k}})$ will then yield the same
$\hat{\rm{g}}(0)$. Showing this is quite straightforward if one
notices that first, the commutator and the anticommutator in Eq.\
(\ref{gf1}) commute with each other, and second,
$\hat{\rm{A}}_1(\rm k)\hat{\rm{A}}_2(\rm
k)=\hat{\rm{B}}_1(\rm{\underline{k}})\hat{\rm{B}}_2(\rm{\underline{k}})=0$.

To evaluate the spin current density at a general position $x$, we
need the traces\cite{Lu3} ${\rm Tr} [\hat \sigma \tau_3 \hat g(u)]$
at $u$. If $\hat d \parallel \hat z$, we see that contributions to
${\rm Tr} [\sigma_z \tau_3 \hat g(u)]$ arises only from the $++$ and
$--$ blocks, whereas ${\rm Tr} [\sigma_x \tau_3 \hat g(u)]$ and
${\rm Tr} [\sigma_y  \hat g(u)]$ arise only from the $+-$ and $-+$
blocks. Hence the latter two have $u$ dependence given by $ e^{ - 2
\alpha_s |u| / v_f}$ (see eq (\ref{eq_out}) and (\ref{eq_in}); note
also ${\rm Tr} [ \hat \sigma \tau_3 \hat g_b ] = 0$).  Hence for
general $\hat d$, if $\hat n$ is a vector perpendicular to $\hat d$,
${\rm Tr} [ (\hat n \cdot \hat \sigma) \tau_3 \hat g(u) ]$ will have
this same $u$ dependence.  Since for our superconductor $\hat d$ is
in the x-y plane, ${\rm Tr} [ \sigma^z \tau_3 \hat g (u)]$ is simply
${\rm Tr} [ \sigma^z \tau_3 \hat g (0)] e^{ - 2 \alpha_s |u| /
v_f}$. The $x$ integral needed for the total spin current is simply
 $\int_0^{\infty} dx e^{ - 2 \alpha_s |u| / v_f} = \frac{ v_f |
{\rm cos} \phi | }{2 \alpha_s}$.

\newpage

\newpage

\begin{thebibliography}{plain}
\bibitem{footnote} We find that the spin polarization obeys $S^2_x+S^2_z=1$ for all
angles except at $\phi=0$, where $S_i = 0$ for all $i$ as a
consequence of the time-reversal invariance of our superconducting
state.
\end{thebibliography}
\end{document}